# Searches for scalars at LHC and interpretation of the findings

Anirban Kundu[1], Alain Le Yaouanc[2], Poulami Mondal[1],

François Richard[2]

[1]Department of Physics, University of Calcutta, 92 Acharya Prafulla Chandra Road, Kolkata 700009, India

[2]Université Paris-Saclay, CNRS/IN2P3, IJCLab[1], 91405 Orsay, France

______________________________________________________________

**Abstract:** *In view of the future Higgs factories, this work presents the status of scalar searches at the LHC with an emphasis on the H(650) resonance which has been observed in WW, ZZ and h(95)h(125) channels, with a cumulative evidence of about 7 s.d. global significance. Its interpretation in models, restricted to extension of the scalar sector by SU(2) singlets and doublets, is clearly excluded, while its interpretation in models with additional triplets requires an extension with respect to the conventional Georgi-Machacek model. A general picture of these searches is updated, showing that h(95) is also reaching a similar level of evidence while two other candidates, A(400) and h(151), although less prominent, are above the 4 s.d. global evidence.*

---

[1] Laboratoire de Physique des 2 Infinis Irène Joliot-Curie

## Introduction

A few words of introduction are needed to understand the motivations which are behind the present work, in the context of the ECFA workshop for future Higgs factories. Before the HEP community can choose between the proposed e+e- colliders, a valid question to ask is: **which energy** is needed to **directly** observe "beyond Standard Model" (BSM) physics? In other words, what are the masses of the lightest BSM particles?

If there are such particles within reach of a TeV e+e- linear collider, they should have already appeared in **LHC present data** as was the case for h(125) at Tevatron. If they don't, there is little hope for a firm discovery in a near future.

Avoiding a « blind search », it is assumed that, as for the Higgs in the SM and the pions in QCD, the lightest new objects are scalars residuals from a **symmetry breaking mechanism** which protect their masses. The next step is to embed these scalars in an extended version of the SM, comprising only additional doublets or singlets, automatically satisfying the $\rho$ **constraint**. At variance with this, we have also considered the **Georgi-Machacek** model [1], GM, which assumes that gauge fields appear in a doublet plus two triplets, one real and the other complex with identical vacuum expectation values (VEV), satisfying the $\rho$ constraint. We are aware that this condition is only satisfied at the tree level, meaning that it is not excluded that this parameter may deviate from one.

Given the short time allocated for this presentation, the topic has been very partially covered, concentrating mainly on the evidence for H(650) in four different modes. During the same session, the case for a lighter candidate labelled h(95) has also been presented [2], with increasing evidence. The reader is referred to previous documents to have a detailed view of our investigations [3]. For what concerns H(650), our previous work was suggesting that H(650) was belonging to H5, a 5-plet of GM, while the recent result from CMS for H(650) → WW in leptonic decays [4] shows that this cannot be the case, suggesting to extend the GM framework, an ongoing task which has not yet been completed.

To be fair, one should also quote other attempts to interpret these observations: [5] and [6].

In the next section are recalled the first evidences for H(650), while section II discusses two recent results which modify our vision of this resonance, specifically that it does not fit into models comprising of only doublets and singlets.

The statistical significances of these signals are updated in section III with a summary of their properties, while section IV summarizes the whole picture. Section V discusses the possible signals at future leptonic colliders, and we end with Section VI which provides a possible theoretical framework to incorporate all these new scalars.

### I. First evidences for a H(650) scalar

At LHC, the cleanest channel to discover new scalar resonances is the 4 lepton channel resulting from H->ZZ. From a combination of published histograms performed in [7], adding 113.5 fb$^{-1}$ from CMS (2/3) and ATLAS (1/3), one observes a peak at $M_H$~650 GeV, with s/b=42/14 ~3.75 s.d. local significance as shown in figure 1. The measured total width is $\Gamma_H$~100 GeV, while the cross section is ~90±25 fb.

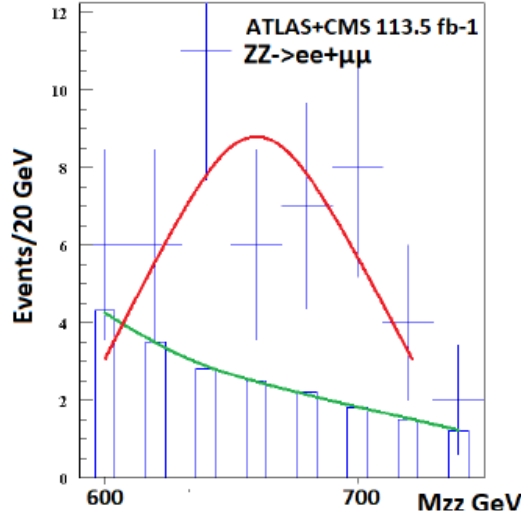

*Figure 1: Evidence for H(650)->ZZ->4leptons combining ATLAS and CMS published histograms [7]*

With 139 fb-1 ATLAS observes a ~3.5 s.d. effect at the same mass [8].

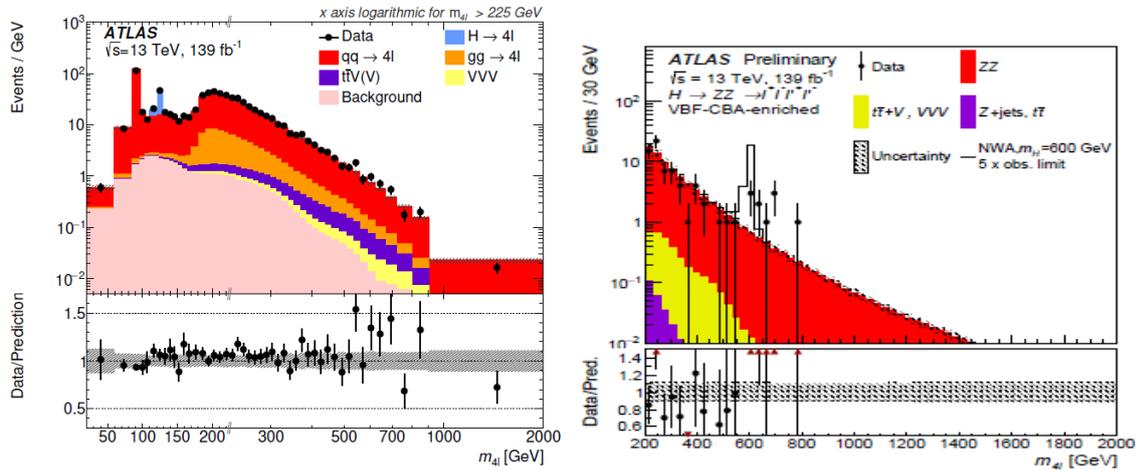

*Figure 2: Inclusive (left diagram) and VBF (right diagram) searches for resonances from ATLAS, using the ZZ channel into four leptons.*

With the same luminosity and using **sequential cuts**, ATLAS also observes an excess at the same mass for the vector boson fusion process VBF, with s/b=9/2 ~2.1 s.d. local significance. The corresponding cross section for VBF->H(650)->ZZ is ~30±15 fb, therefore significantly below the inclusive one, implying a **ggF contribution** of 60 fb. This, by itself suggest that H(650) is not a pure H5 custodial 5-plet state, since it couples to fermions, a prerequisite to generate a ggF contribution.

## II. Recent evidences for H(650)

### II.1 pp(VBF) → H(650) → WW

CMS has produced new evidence for H(650) in the VBF process VBF → H(650) → WW → ℓℓνν, with 3.8 s.d. local significance, a cross section ~160±50 fb, close to the cross section predicted for a SM Higgs model [3]. This cross section is 5 times larger than for VBF → H(650) → ZZ, therefore **inconsistent with the H5 interpretation,** which predicts cross-sections for WW/ZZ=0.5. This means that the canonical GM model is unable to interpret H(650).

The following table indicates the various scenarios envisaged to interpret the CMS data where one sees that the largest excess is observed for a mass of 650 GeV for a VBF cross section of ~160 fb.

Table 3: Summary of the signal hypotheses with highest local significance for each $f_{VBF}$ scenario. For each signal hypothesis the resonance mass, production cross sections, and the local and global significances are given.

| Scenario | Mass [GeV] | ggF cross sec. [pb] | VBF cross sec. [pb] | Local signi. [$\sigma$] | Global signi. [$\sigma$] |
|---|---|---|---|---|---|
| SM $f_{VBF}$ | 800 | 0.16 | 0.057 | 3.2 | 1.7 ± 0.2 |
| $f_{VBF}=1$ | 650 | 0.0 | 0.16 | 3.8 | 2.6 ± 0.2 |
| $f_{VBF}=0$ | 950 | 0.19 | 0.0 | 2.6 | 0.4 ± 0.6 |
| floating $f_{VBF}$ | 650 | $2.9 \times 10^{-6}$ | 0.16 | 3.8 | 2.4 ± 0.2 |

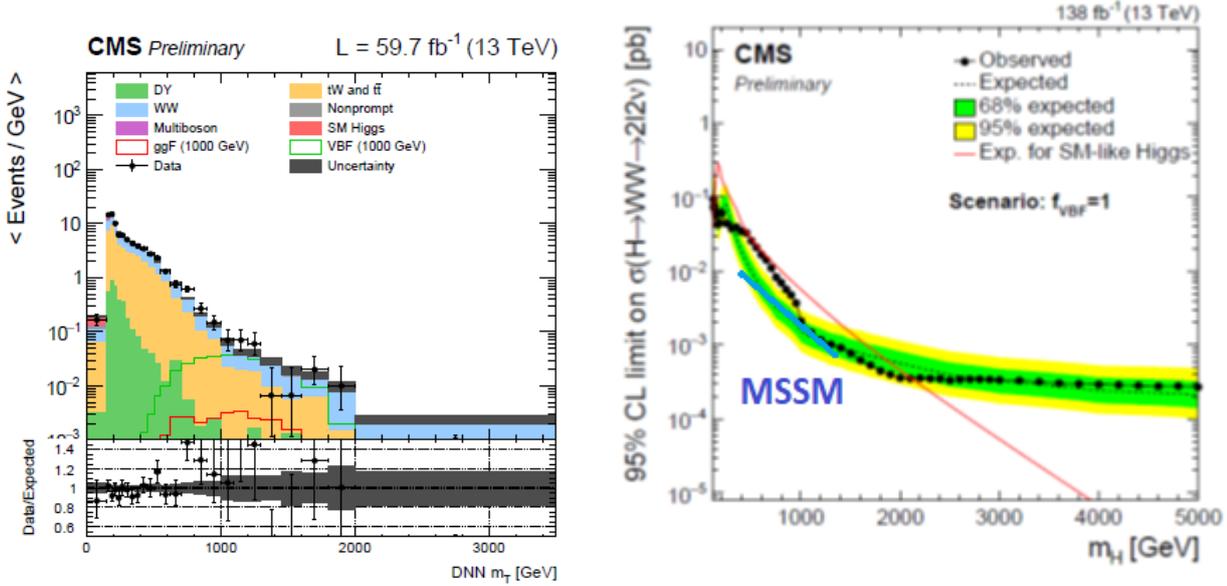

*Figure 3: CMS search for resonances decaying into WW->ℓℓ νν, produced in the VBF mode. The right panel shows an excess which reaches the SM prediction (red line) for a heavy Higgs, while the blue line shows an upper limit for an MSSM prediction, taking into account the CMS measurement of h(125)->WW.*

Within two doublet models, one can relate the H(650)->WW to h(125)->WW. Taking a recent result of CMS [9], in the standard notation for two doublet models, one predicts:

$$\sin^2(\alpha-\beta) \sim 0.97 \pm 0.09$$

meaning that the coupling H(650)WW should be suppressed by $\cos^2(\alpha-\beta) \sim 0.03 \pm 0.09$, in total contradiction with the CMS result where this coupling is close to 1. Figure 3, right histogram, summarizes this situation.

This exclusion of two doublet models results from general sum rules and we shall see that this does not apply to models with triplets where doubly charged scalars contribute with opposite sign to these sum rules and introduce a compensating term which allows the large coupling observed by CMS for H(650)->WW (see section VI and reference [19]).

## II.2 H(650) ->bbh(125)->bbγγ

At the summer conference ICHEP22, CMS has delivered another striking effect around 650 GeV, in the process [10]:

$$H->bbh(125)->bb\gamma\gamma$$

showing an excess for mH=650 GeV for a bb mass around 90 GeV, with a 3.8 s.d. local significance.

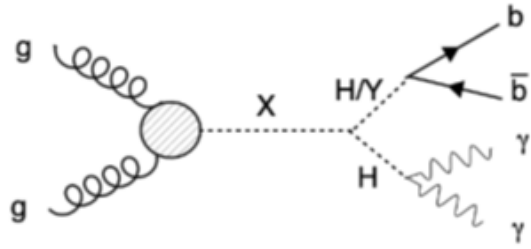

The poor mass resolution achieved in reconstructing the bb excess does not allow to unambiguously distinguish between h(95)->bb and Z->bb, but the final state Zh(125) would be inconsistent with a CP-even H(650) decaying into ZZ/WW.

This excess corresponds to ~0.4 fb. Taking into account the BR of h(125) into two photons and assuming that h(95) decays into bb in ~90% of the cases, one gets a cross section ~200±60 fb. Since there is no selection for VBF, this cross section includes both VBF and ggF.

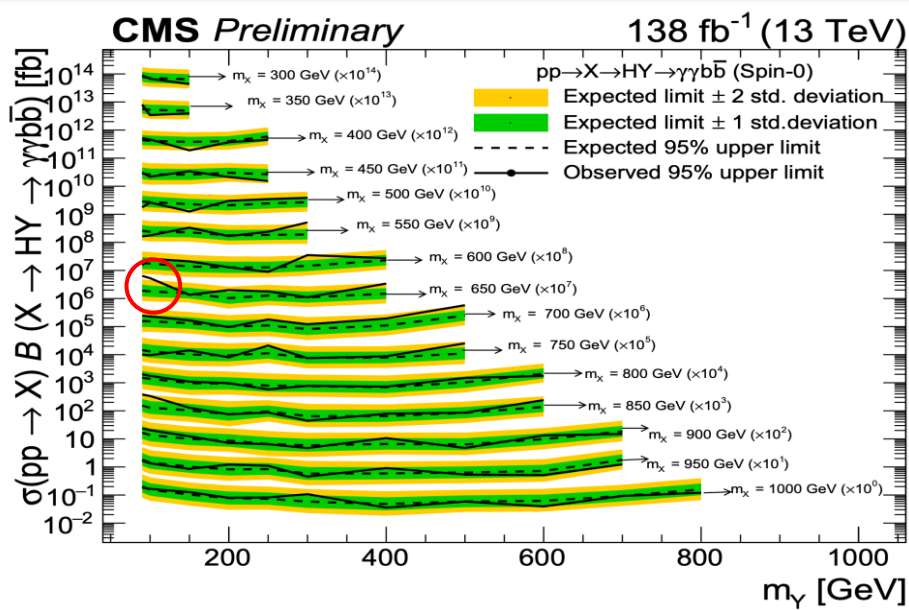

*Figure 4: Cross section for pp->X->h(125)Y selecting h(125) into two photons and Y into bb, versus mY, for various selections of mX. Each line gives an 95% upper limit for a given mass of X. For mX=650 GeV, in the region with a red circle at mY~90 GeV, an excess is observed.*

One eagerly awaits for a similar analysis from ATLAS before drawing a firmer conclusion.

## II.3 One or two resonances?

From the measured VBF cross sections into ZZ and WW, one may infer the two partial widths $\Gamma_{ZZ}$=15±8 GeV and $\Gamma_{WW}$=75±20 GeV. Adding these two already gives a width comparable to the measured total width in the four lepton mode.

The inclusive measurement ggF+VBF into ZZ also suggests that ggF ~60±30 fb, meaning that one should add a partial width into top pairs of order 15 GeV. This comes from the SU(2) doublet component of H(650).

Finally the process H(650)->h(95)h(125) contributes to a cross section of 200±60 fb and should therefore give a partial width of the same order as WW, in tension with the measured total width in the ZZ mode.

Admittedly, all these measurements have large errors which prevents a definite statement but they suggests that in interpreting H(650), one cannot exclude that there could be two distinct resonances with similar masses, which would solve the apparent width paradox. The two resonances need not be close to being degenerate, as the width is large. The occurrence of such close-by resonances is not unlikely in a theory with many scalar fields, like the extended GM model discussed in section VI, which would include such additional scalars.

## III. Statistical significance of these signals

The table below summarizes the significances of the various indications reported in our previous ArXiv documents, with some updates commented below.

A general comment is of order. When combining several indications, one assumes that they originate from the **same resonance** which, given the poor mass resolution, cannot be firmly proven. In the case of H(650), the total width of this resonance, as seen in 4 leptons is ~100 GeV, meaning that there is a finite probability of coincidence with an unrelated resonance, e.g. H'->h(125)h(95), which would not contribute to the total width of H(650). In such a case the significance of the signal should be decreased.

| Reaction | # channels/expts | # σ global (loc) | Michelin rating |
|---|---|---|---|
| pp->h(125) | >2/2 | 6.9 | *** |
| pp->X(750) | 1/2 | 4.3 (dead) | * |
| pp->A(400) | 3/2 | 5 | * |
| pp->H(650) | 4/2 | 7.5 | ** |
| pp->SSℓ,ℓ+b,… | >2/2 | 8 | * |
| pp->H(151)+Z | 1/2 | 4.8 | * |
| h(95) LHC+LEP2 | 3/2 | 4.3 | * |
| pp->H5+(350)->WZ | 1/2 | 3.5 | |
| h(125)->a(52)a(52) | 1/1 | 1.7 (3.3) | |
| pp->H3+(130)->bc | 1/1 | 1.6 | |

### III.1 Status of A(400)

For what concerns A(400) [11], it is fair to say that the progress of confirmation of this signal is rather slow, with ups and downs. Since CMS has not completed its RUN2 analysis for top pairs nor ATLAS joined in for this challenging work, one remains at the initial 3.5 s.d. local significance. Other evidence, like A->Zh and A->ττ with or without an accompanying b jet, which were based on a partial analysis of RUN2 data, have not been confirmed. ATLAS has recently produced an impressive new result [12], which claims observing A->ZH(320)->Zh(125)h(125), as shown in figure 5. In passing, this would be the second evidence for a triple Higgs coupling in an unexpected scenario.

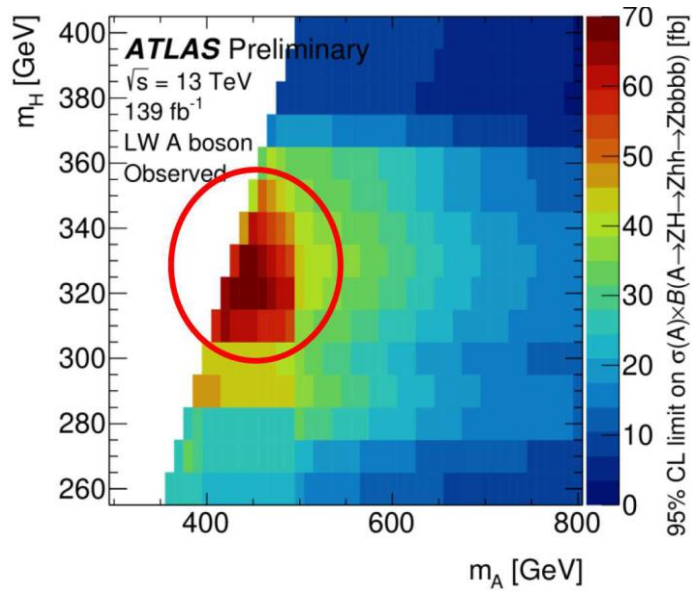

*Figure 5: ATLAS search for A(400)->ZH->Zhh.*

### III.2 Status of h(95)

For what concerns h(95), there is an obvious progress with the appearance of the cascade:

$$H(650) \to h(95)h(125) \to bb\gamma\gamma$$

As already noted however, this channel does not necessarily belong to the same resonance decaying into ZZ/WW and being conservative, we ignore, for the moment, this additional significance.

### III.3 Status of h(151)

An indirect proof of the existence of A(400) comes from a plausible interpretation of the evidence for h(151)->γγ+ETmiss as coming from A->h(151)+Z, with Z into neutrinos [13]. To be on the safe side, this evidence will not be considered for the significance of A. If it were, A would also reach ~7 s.d. global significance.

At this moment, the combined evidence for A(400) is at a the level of 5 s.d. global significance.

## IV. The emerging picture

One can summarize the strongest LHC findings in the diagram below.

Most of these legends are self-explanatory with some exceptions: the **aa mode**, which is added for h(125) and h(151) refers to the indication found by ATLAS [14] for the decay of h(125) into a pair or axial scalars, with m(a)=52 GeV, in the mode bbμμ. In the absence of such a contribution, h(151), assumed to be an isosinglet, would dominantly decay into ZZ/WW. In the cascade interpretation A->h(151)Z, one would end up with an unacceptably large amount of ZWW events, which is excluded by present searches [13]. A way out was therefore to assume that h(151) decays predominantly into aa. Recently [15] CMS has searched for h(125)->aa->bbµµ and, with a better sensitivity, has found no evidence for this process without being able to exclude its presence.

Note also that the evidence for h(151) fulfils the 4 s.d. criterion but results from an unofficial combination of ATLAS and CMS data.

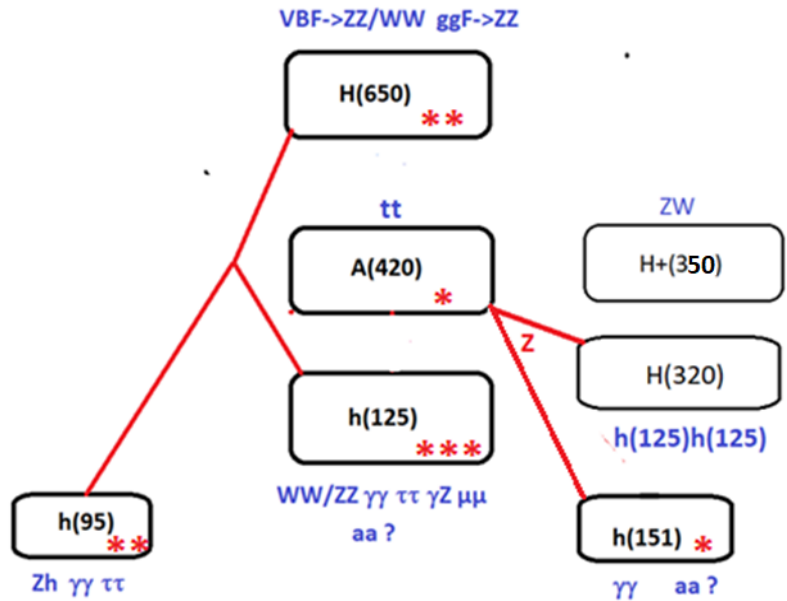

The presence of a charged scalar H+(350) decaying into ZW is still marginally significant. It has been seen by both experiments but is not yet crossing the 4 s.d. global significance required to join the club. This is also true for H(320) seen into h(125)h(125) through the cascade A->H(320)Z->hhZ.

ATLAS [17] has attempted to combine the searches for H5 into ZZ->ℓ+ℓ-ℓ'+ℓ'-, H+->ZW+ and H++->W+W+ , assuming a common mass m5 for these particles and varying the mixing parameter sin($\theta_H$), hereafter abbreviated by sH. They have also searched for the DY process:

pp->H5++ H5- - ->W+W+W-W-

which allows to eliminate the dependence on sH.

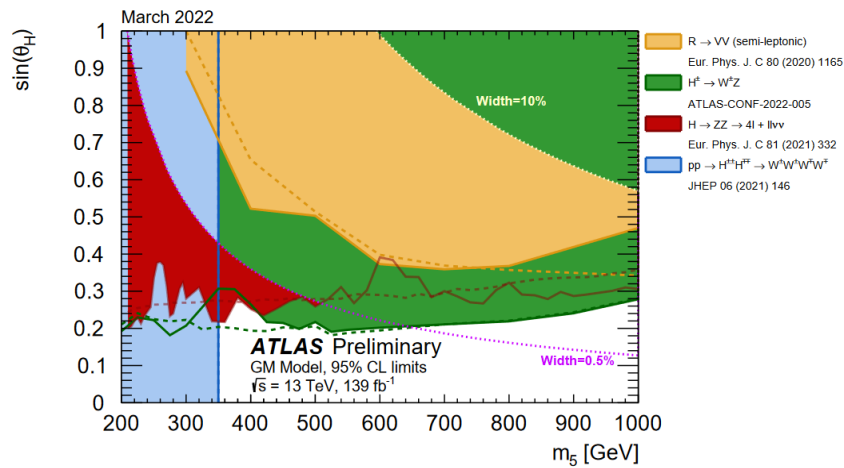

*Figure 6: The reader is referred to [17] for detailed explanations of these exclusion contours, which are based on ZZ, ZW and same sign WW leptonic decay modes. The blue band exclusion is based on the DY process pp->H++H - - ->W+W+W-W- searches.*

Figure 6 shows an impressive coverage reached by these searches, suggesting that sH<0.25 in the whole relevant mass domain. These exclusions however do not consider the potential cascades of these particles into the triplet H3+ and A, like for instance H5++ ->H3+W+. This also applies to H5 and H5+ which could decay into H3+W- or H3+Z. These cascades, not suppressed by the mixing sH, will tend to dominate, weakening the limit on sH derived from ZZ, W+W+ and ZW+ modes assuming a branching ratio of 100%.

There is a candidate CP-odd object A with a mass ~400 GeV, which can be part of a custodial triplet. If one makes the usual assumption that the triplet and 5-plet scalars are degenerate in mass, this implies that such cascades will only be opened for m5>480 GeV. This would imply that for m5<480 GeV, the ATLAS picture holds. Reference [17] however tells us that one can relax this mass degeneracy, meaning that this 480 GeV limit on m5 is model dependent.

The lower mass limit 350 GeV set on mH++ does not depend on sH since it relies on DY pair production of these objects decaying into W+W+W-W-. It is should be true unless H3+ is lighter than 350-mW=170 GeV. This small value seems unlikely given indirect constraints provided by b->sγ but this statement is model dependent. In this respect, one may quote a weak indication at mH+=130 GeV for H+->bc found in [18] at 3 s.d. level.

Again, as pointed out in [17], mH++ >350 GeV does not signify that mH5>350 GeV nor mH5+>350 GeV. Excluding a light H5 is non-trivial since, with sH~0.2, its production cross section is small of order 10 fb. The experimental width of ZZ is dominated by the experimental resolution, which means that this search is optimal in the 4 leptons mode. The sensitivity reached by ATLAS is not yet sufficient to exclude this low mass scenario.

In summary, one may conclude that the absence of indications for the H5 scalars could either be due to a very small value of sH, below 0.3, which suppresses the WW/ZZ/ZW channels or due to the early opening of cascades with a light triplet H3. The indication of H5+->ZW with mH5+~350 GeV, if confirmed, would exclude a very light H3. It would also set a minimum value for sH, sH>0.2, and therefore indicate the sensitivity needed to observe the gold-plated signal H5++, unless mH5++ is much heavier than mH5+.

Clearly the final proof of GM will rely on finding H5+->ZW or, even better, H5++->W+W+.

## V.     Predicted cross sections in e+e-

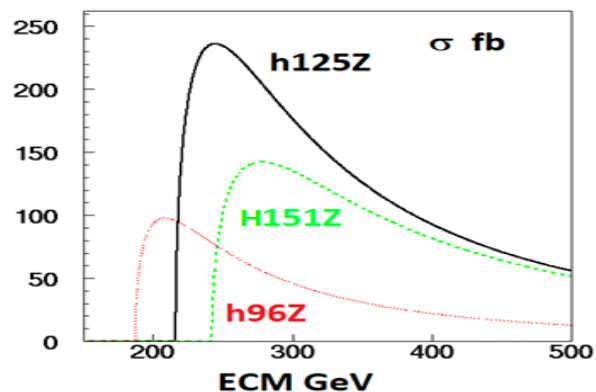

*Figure 7: Predicted cross sections for the lightest scalars versus the center of mass energy.*

One can distinguish two populations. The SM Higgs boson and the two light Higgs candidates h(95) and h(151), with copious cross sections, as shown in the figure 7. All presently considered Higgs factories would be able to produce them abundantly and achieve similar high accuracies as for h(125).

The second category is more challenging and requires achieving very high luminosities and reaching no less than 1 TeV centre of mass energy. It corresponds to a very rich landscape but with cross sections much smaller, as shown in figure 8. A special case concerns H(650) which is best produced by VBF as suggested by the result of CMS.

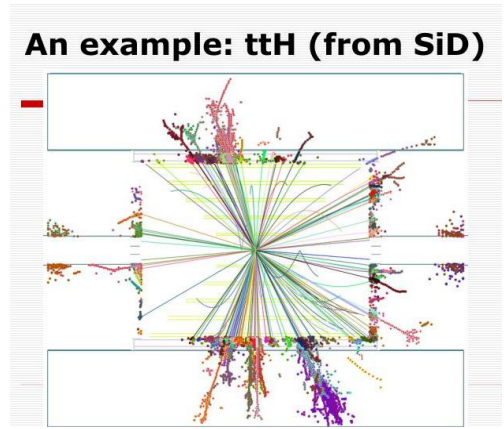

In the ILC scenario, one will be able to collect 8000 fb-1 at 1 TeV, which is barely sufficient and requires an excellent detector able to record the complex modes which will be produced at 1 TeV. This is already the case for the SM final state tth(125), as shown in above picture, where 8 jets are produced, meaning that the reconstruction efficiency goes like $\Omega^8$, where $\Omega$ is the solid angle covered for reconstructing jets. Solenoidal detectors tend to have a poor detection efficiency in the forward regions, both for particle identification and for b/c tagging properties, which does not hamper physics at low energy but could be fatal at 1 TeV.

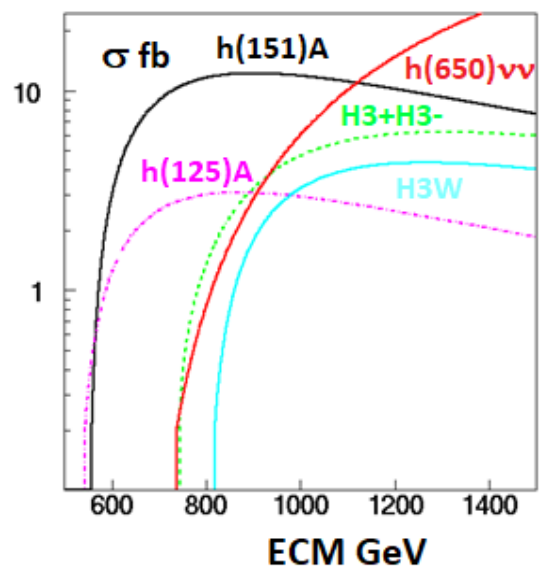

*Figure 8: Predicted cross sections for the heavy scalars versus the centre of mass energy. Note that H(650) is best produced by the VBF process.*

In the GM scenario, it will be critical to study doubly charged Higgses. They can be pair produced but this may require rising the energy well above 1 TeV. An alternate possibility would be to run the collider in an **e- e- mode**, using the VBF process to produce H- -. The expected cross section is shown below assuming a mass of 650 GeV for H- -.

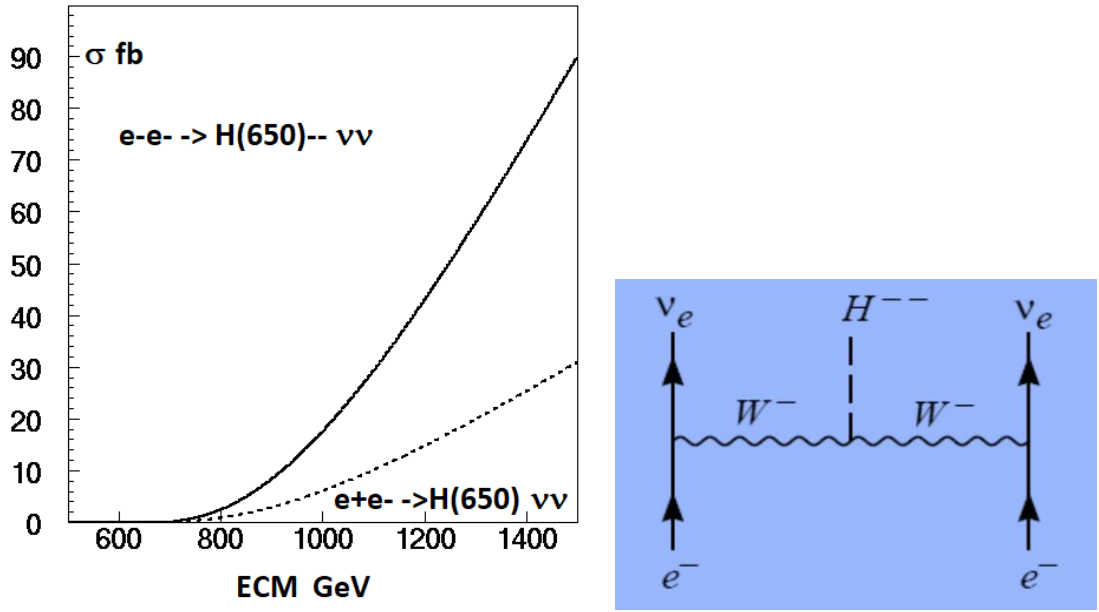

Figure 9: VBF cross sections for e+e-->H(650) νν and for e-e-->H - - νν.

## VI. An extended GM model eGM?

Similarly to the original two-Higgs doublet models (2HDM), GM can be extended in many ways by adding singlets and doublets. This applies also to the pattern of Yukawa couplings where there are large possibilities exploited by phenomenology. We will not talk about the extension of the GM model with the same number of degrees of freedom but less constraints on the potential [15].

The evolution of GM will be dictated by the rich data coming from LHC and in our previous notes, on various occasion, this need was expressed for explaining some indications of A->ττ or the presence of b jets accompanying Zh. One may note that the large VBF cross-section for H(650) → WW rules out 2HDM as a possible solution, as the unitarity sum rule is violated without a doubly charged Higgs [19].

The interpretation of H(650) → ZZ/WW clearly dictates the need to go beyond the canonical GM, after discarding the H5 interpretation. Recall the two main reasons for this:

- The VBF process produces a much larger rate into WW than into ZZ, while the H5 decay predicts a ratio ZZ/WW=2
- The inclusive production of ZZ is larger than the VBF cross section, while H5 is expected to be produced solely by VBF and not by ggF

For a plausible solution, we propose an extension of the GM model with another SU(2) doublet Φ2, in addition to the usual doublet and two triplets. There are now three independent VEVs and so there must be three custodial singlets. Indeed, the model respects custodial symmetry and the final custodial multiplets are

$$5 + 3 + 3 + 3 + 1 + 1 + 1$$

where one of the triplets is the Goldstone. There are 4 CP-even scalars, namely, three custodial singlets and the neutral of the 5-plet that does not develop any VEV. We can identify the three singlets with h(95), h(125), and H(650) respectively and assume that X(151) and A(400) reside in the remaining two CP-odd triplets. Note however that this interpretation of X(151) is incompatible with the cascade assumption A->ZX(151) interpreting the tagging properties of this channel [13]. The exact nature of the parameter space, the constraints, and possible signals, are under investigation and will soon be reported. Note that this model can also accommodate the newcomer H(320), not yet reaching the canonical significance, but which cannot be accommodated in the standard GM.

## Summary and conclusions

Recent findings have reinforced the evidence for H(650) and h(95). **H(650)**, first indicated in the **ZZ** mode by ATLAS is now observed by CMS into **WW** through VBF production and in the triple Higgs process **H(650)-> h(95)h(125).**

For H(650)->WW, the VBF process gives a large cross section close to the SM, clearly incompatible with standard models with isodoublets and isosinglets.

Comparing the VBF processes for ZZ and WW, one reaches the conclusion that a 5-plet solution H5 from GM is also excluded, meaning that H(650) does not find its place within the genuine GM model.

The recent observation of the process A(420)->ZH(320)->Zhh nicely confirms the CP-odd resonance but cannot find its place in the genuine GM model.

h(95), h(151), H(320) and H(650) could be interpreted as isosinglets, while the GM model only allows for one isosinglet.

Extending the GM model with additional isosinglets, isodoublet(s) and/or isotriplet(s) pauses no basic problem and should solve most of these puzzles. The issue for H(650)->ZZ/WW seems the most arduous to resolve with simple-minded additions. A more subtle approach is under way which assumes that H(650) could be a complex mixture of states provoking a destructive interference for the ZZ final state, constructive for WW.

In any case, these observations offer a promising new landscape for HEP, in particular for future e+e- colliders under discussion and motivate a **linear e+e- collider reaching no less than 1 TeV** in order to produce these various states.

Complex final states are generated by cascade processes of the type e+e-->AH(320)->tthh or Zhhhh will have a **critical impact in the design of future LC detectors, requiring a full acceptance coverage to reconstruct multijet events.**

**Acknowledgment**s: We are very grateful to Gilbert Moultaka for many useful comments on this work. It is also a pleasure to acknowledge useful discussions with Howard Haber and Sven Heinemeyer.